\documentclass[twocolumn,prl,showpacs,preprintnumbers,superscriptaddress]{revtex4}

\usepackage{color}
\usepackage{graphicx}
\usepackage{dcolumn}
\usepackage{bm}
\usepackage{verbatim}
\usepackage{amsmath}

\newcommand{\ket}[1]{| #1 \rangle}
\newcommand{\bra}[1]{\langle #1 |}

\usepackage[hypertex]{hyperref}
\newcommand{\nc}{\newcommand}
\nc{\be}{\begin{eqnarray}}
\nc{\ee}{\end{eqnarray}}
\nc{\bea}{\begin{eqnarray}}
\nc{\eea}{\end{eqnarray}}
\nc{\bean}{\begin{eqnarray*}}
\nc{\eean}{\end{eqnarray*}}
\nc{\mb}{\mbox}
\nc{\rnc}{\renewcommand}
\nc{\vk}{\mb{\bf k}}
\nc{\vp}{\mb{\boldmath$p$}}
\nc{\rr}{\mb{\boldmath$r$}}
\nc{\vR}{\mb{\boldmath$R$}}
\nc{\vz}{\hat {\mb{\bf z}}}
\nc{\vj}{\mb{\boldmath$j$}}
\nc{\vg}{\mb{\boldmath$g$}}
\nc{\vE}{\mb{\boldmath$E$}}
\nc{\vB}{\mb{\boldmath$B$}}
\nc{\vH}{\mb{\boldmath$H$}}
\nc{\vM}{\mb{\boldmath$M$}}
\nc{\vP}{\mb{\boldmath$P$}}
\nc{\vS}{\mb{\boldmath$S$}}
\nc{\x}{\mb{\boldmath$x$}}
\nc{\A}{\mb{\boldmath$A$}}
\nc{\va}{\mb{\boldmath$a$}}
\nc{\vq}{\mb{\boldmath$q$}}
\nc{\vn}{\mb{\boldmath$n$}}
\nc{\vs}{\mb{\boldmath$\sigma$}}
\nc{\vt}{\mb{\boldmath$\tau$}}
\nc{\vpi}{\mb{\boldmath$\pi$}}
\nc{\nab}{\bm{\nabla}}
\nc{\X}{\sf x}
\renewcommand{\vec}[1]{\mathbf{#1}}

\begin{document}

\title{
Giant magneto-optical response in non-magnetic semiconductor BiTeI driven by bulk Rashba spin splitting}

\author{L. Demk\'o}
\affiliation{Multiferroics Project, Exploratory Research for Advanced Technology (ERATO),
Japan Science and Technology Agency (JST), c/o Department of Applied Physics,
University of Tokyo, Tokyo 113-8656, Japan}

\author{G. A. H. Schober}
\affiliation{Department of Applied Physics, University of Tokyo, Tokyo 113-8656, Japan}
\affiliation{Institute for Theoretical Physics, University of Heidelberg, D-69120 Heidelberg, Germany}

\author{V. Kocsis}
\affiliation{Department of Physics, Budapest University of
Technology and Economics and Condensed Matter Research
Group of the Hungarian Academy of Sciences, 1111 Budapest, Hungary}

\author{M. S. Bahramy}
\affiliation{Cross-Correlated Materials Research Group (CMRG) and
Correlated Electron Research Group (CERG), RIKEN Advanced Science Institute (ASI),
Wako 351-0198, Japan}

\author{H. Murakawa}
\affiliation{Cross-Correlated Materials Research Group (CMRG) and
Correlated Electron Research Group (CERG), RIKEN Advanced Science Institute (ASI),
Wako 351-0198, Japan}

\author{J. S. Lee}
\affiliation{Department of Applied Physics, University of Tokyo, Tokyo 113-8656, Japan}

\author{I. K\'ezsm\'arki}
\affiliation{Department of Physics, Budapest University of
Technology and Economics and Condensed Matter Research Group of
the Hungarian Academy of Sciences, 1111 Budapest, Hungary}

\author{R. Arita}
\affiliation{Department of Applied Physics, University of Tokyo, Tokyo 113-8656, Japan}
\affiliation{Cross-Correlated Materials Research Group (CMRG) and
Correlated Electron Research Group (CERG), RIKEN Advanced Science Institute (ASI), Wako 351-0198, Japan}

\author{N. Nagaosa}
\affiliation{Department of Applied Physics, University of Tokyo, Tokyo 113-8656, Japan}
\affiliation{Cross-Correlated Materials Research Group (CMRG) and
Correlated Electron Research Group (CERG), RIKEN Advanced Science Institute (ASI), Wako 351-0198, Japan}

\author{Y. Tokura}
\affiliation{Multiferroics Project, Exploratory Research for Advanced Technology (ERATO),
Japan Science and Technology Agency (JST),
c/o Department of Applied Physics, University of Tokyo, Tokyo 113-8656, Japan}
\affiliation{Department of Applied Physics, University of Tokyo, Tokyo 113-8656, Japan}
\affiliation{Cross-Correlated Materials Research Group (CMRG) and
Correlated Electron Research Group (CERG), RIKEN Advanced Science Institute (ASI),
Wako 351-0198, Japan}

\date{\today}

\begin{abstract}
We study the magneto-optical (MO) response of polar semiconductor BiTeI with giant {\it bulk} Rashba spin splitting at various carrier densities.
Despite being non-magnetic, the material is found to yield a huge MO activity
in the infrared region under moderate magnetic fields ($\leq 3$\,T).
By comparison with first-principles calculations, we show that such an enhanced MO response is mainly due to the
intraband transitions between the Rashba-split bulk
conduction bands in BiTeI, which give rise to distinct novel features
and systematic doping dependence of the MO spectra.
We further predict an even more pronounced enhancement in the low-energy MO response and dc Hall effect
near the crossing (Dirac) point of the conduction bands.
\end{abstract}

\maketitle

The spin-orbit interaction (SOI) plays a crucial role in the rapidly evolving field of spintronics
\cite{awschalom,parkin,sinova2,winkler}. The principal importance of SOI is in its ability to intrinsically couple the electron spin with its orbital motion, and hence produce many novel phenomena such as the spin Hall effect \cite{murakami} and spin Galvanic effect~\cite{ganicgev}. In the presence of an external magnetic field, SOI can effectively mediate the interaction between photons and electron spins, thereby leading to interesting  magneto-optical (MO) effects, e.g. non-linear Kerr rotation~\cite{bennemann} whereby the polarization plane of the linearly-polarized light in reflection is rotated as a consequence of SOI.

In practice, most materials under magnetic field exhibit a rather complicated MO response. This is because usually too many energy bands are involved in the optical excitations, which, given that SOI is also at work, leads the respective inter- and intraband optical transitions to produce complex spectra. This accordingly prevents a comprehensive understanding of the role of SOI on the MO response of the given materials using the available theoretical models.
In contrast, semiconductors with Rashba-split conduction/valence bands appear to be ideal systems for studying MO effects, as they have a rather simple  spin-dependent multi-band scheme. However, these systems
have rarely been studied up to now, mainly because they usually show a very weak Rashba spin splitting (RSS) which cannot be resolved experimentally, and also because RSS had been found only in the two-dimensional electron-gas or metallic systems
formed at the surface or interface where the MO effect is hardly detectable.

This situation has been greatly improved recently by the discovery of the giant {\it bulk}
RSS in the polar semiconductor BiTeI. The angle-resolved photoemission spectroscopy (ARPES) \cite{bitei_arpes} has revealed that the bulk conduction bands in BiTeI are subject to a large RSS (see Fig.~1(a)), well describable by the 2D Rashba Hamiltonian
$H_R = {{ \vec p^2 / 2 m}} + \lambda {\vec e}_z \cdot ( {\vec S} \times {\vec p})$,
near the time-reversal symmetry point A, where ${\vec e}_z$ is the direction of the potential gradient breaking the inversion symmetry,
and ${\vec S}$ and ${\vec p}$ are the spin and momentum operators, respectively.
This leads to a Dirac-like band dispersion near ${\vec p}={\vec 0}$
and a splitting increasing linearly with $|{\vec p}|$. The succeeding first-principles calculations~\cite{bitei_theory,bitei_pressure} and the optical conductivity spectra~\cite{bitei_sigmaxx} have been consistent with the APRES data and further revealed that not only the bottom conduction bands (BCBs) but also the top valence bands (TVBs) are subject to a comparable RSS in BiTeI. This condition allows several distinct inter- and intraband transitions between these two sets of states. The reduced dimensionality together with RSS
introduces a well-defined singularity in the joint density of states at the band edge.
Therefore, this material is a promising candidate to host the enhanced spin-charge coupling
and/or the magnetoelectric effect with the possible
applications to spintronics. From this viewpoint, the MO effect is an
intriguing issue in BiTeI; the novel interplay between RSS and
external magnetic field has been extensively investigated from theoretical side \cite{shen, burkov, xu, kushwaha}
but hardly ever probed experimentally in a real compound.
Of particular interest are the states near the band crossing (Dirac) point formed by RSS, as they turn into novel Landau
levels under magnetic field similar to the case of graphene \cite{shen}. Thus, it is important to know their impact
on the MO response, e.g. by tuning the Fermi level $E_F$ across the Dirac point. In this Letter, we accordingly study in detail the MO properties of BiTeI both experimentally and theoretically. It is shown that due to giant RSS of bulk bands, the material exhibits a huge enhancement in its MO response in the infrared spectra under a moderately low magnetic field of 3~T. This enhancement is further anticipated to be even more pronounced in the low-energy region near the Dirac point of BCBs.

BiTeI has a trigonal crystal structure with $C_{3v}$ symmetry composed of consecutive Bi, Te, and I layers stacking along the $c$-axis of the crystal \cite{bitei_arpes,bitei_theory}. Due to ionicity and covalency of Bi-I and Bi-Te bonds, the bulk material possesses an intrinsic polarity along the $c$-axis which, when coupled with the strong atomic SOI of Bi, leads to a huge bulk RSS. The calculated band structure of BCBs and TVBs around the band gap (along $H-A-L$ direction), together with the possible inter- and intraband transitions, is shown in Fig.~1(a).
\begin{figure}[ht!]
\includegraphics[width=3.4in]{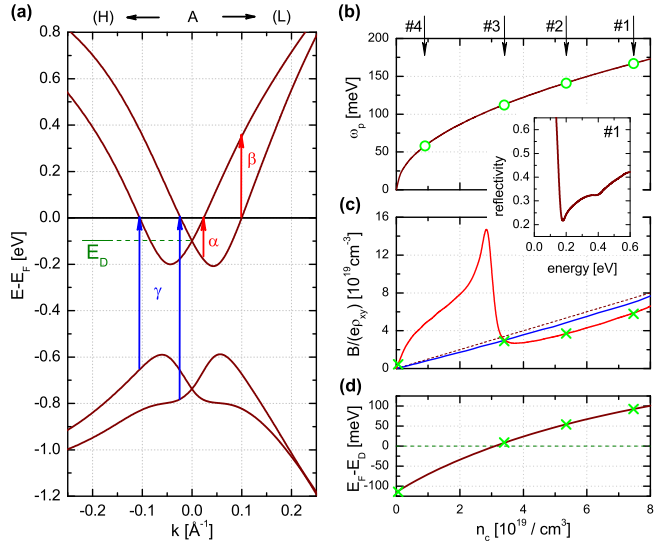}
\caption{(Color online) (a) Calculated electronic band structure of BiTeI with the $\alpha$, $\beta$ (intraband), and $\gamma$ (interband) transitions for a given Fermi level $E_F$. $E_D$ depicts the position of the Dirac point. (b)-(c) carrier density determined from the plasma frequency (panel (b), circles) and Hall response (panel (c), crosses, see the related discussion in the text). Vertical arrows indicate the samples investigated experimentally, while red, blue, and dashed curves of panel (c) represent the calculated Hall response with/without taking into account SOI, and the conventional relation $R_H=-1/(n_ce)$, respectively.
The inset of panel (b) shows the measured low-temperature reflectivity of sample \#1. (d) Fermi level as a function of the carrier concentration $n_c$.}
\end{figure}

The MO effects are discussed by analyzing the elements of the optical conductivity tensor. For the determination of the
off-diagonal components, first the reflectivity spectra were measured over a broad energy
range ($E=10$~meV$-40$~eV and $E=10$~meV$-6$~eV at room and low temperatures, respectively); for measurements above $6$~eV, the synchrotron radiation from UVSOR at the Institute for Molecular Science, Okazaki, was utilized.
The diagonal part $\sigma_{xx}$ has been derived through the Kramers-Kronig transformation.
As the second step, the complex magneto-optical Kerr angle $\Phi_K$ was measured by a
magneto-optical Kerr effect (MOKE) setup: The combination of a Fourier-transform infrared
spectrometer, a polarization modulation technique \cite{moke} afforded by a ZnSe photoelastic
modulator, and a room temperature-bore magnet has enabled the direct derivation of the
off-diagonal conductivity $\sigma_{xy}$ via the following relation:
\begin{equation}
\Phi_K=\theta_K+\rm i\eta_K=-\frac{\sigma_{xy}}{\sigma_{xx}\sqrt{1+(4\pi\rm i/\omega)\sigma_{xx}}}.
\label{eq:Kerr}
\end{equation}
The applicable energy range of the setup is $80-550$~meV.
The experiments were carried out in reflection configuration with nearly normal
incidence on cleaved surfaces of the $ab$-plane. The low-temperature measurements were
performed at $T=10$~K, while the external magnetic field of $B=\pm3$~T has been applied
perpendicular to the sample surface in the case of the MOKE measurements.

To analyze the respective MOKE data, the longitudinal conductivity $\sigma_{xx}$ has been derived from the Kubo formula as described in \cite{bitei_sigmaxx},
while the transverse conductivity $\sigma_{xy}$ has been calculated to linear order in the external magnetic field $B$
using the Fukuyama formula~\cite{fukuyama},
\begin{align}
 & \sigma_{xy}(i\omega_{\ell}) = B \frac{e^3 \hbar}{2\omega} k_B T \sum_n \frac{1}{V} \sum_{\vec k} \nonumber \\
 & \  \Big\{ \, m \, \mathrm{Tr} \Big[ G_- v_x G v_x G - G_- v_x G_- v_x G \Big] \nonumber \\
 & \  \phantom{\Big\{} + m^4 \, \mathrm{Tr} \Big[ G_- v_y G_- v_x G v_x G v_y - G_- v_x G_- v_x G v_y G v_y \nonumber \\
 & \  \phantom{\Big\{} \quad + v_x G v_y G_- v_x G_- v_y G_- - v_x G v_y G_- v_y G_- v_x G_- \nonumber \\
 & \  \phantom{\Big\{} \quad + G_- v_x G v_x G v_y G v_y - G_- v_x G v_y G v_x G v_y \Big] \label{eq:Fukuyama}
   \, \Big\},
\end{align}
where $G = [i\tilde \epsilon_n + E_F - H]^{-1}$ is the one-particle thermal Green's function,
$\tilde \epsilon_n = \epsilon_n + \frac{\Gamma}{2} \mathrm{sgn}(\epsilon_n)$ with $\epsilon_n = (2n+1)\pi k_B T$,
\,$\Gamma$ the damping constant, and $G_-$ is obtained from $G$ by replacing
$\epsilon_n \to \epsilon_n + \omega_{\ell}$ with $\omega_{\ell} = 2 {\ell} \pi k_B T$; $m$ denotes
the electron mass, $V$ the volume of the system, and $v_i$ ($i = x, y$) the velocity operator
represented in matrix form~as
\begin{equation}
        v_{i,nm} = \bra n v_i({\mathbf k}) \ket m = \frac{1}{\hbar} \Big\langle
n \, \Big| \, \frac{\partial H({\mathbf k})}{\partial k_i} \, \Big| \, m
\Big\rangle,
\end{equation}
where $\ket n$ corresponds to the $n$th eigenstate of $ H(\mathbf k)$. These matrix elements were computed
by downfolding the \textit{ab-initio} Hamiltonian~\cite{ab-initio} to a low-energy tight binding model, which was
constructed for the 12 valence and 6 conduction bands around the band gap \cite{bitei_theory}.
To evaluate the formula~\eqref{eq:Fukuyama}, the sum over Matsubara frequencies was transformed
into a contour integral along the branch cuts of the Green's functions~\cite{mahan}. After performing
the analytic continuation $i\omega_{\ell} \to \omega$, the contour integral as well as the
$\vec k$ momentum integral were evaluated numerically. In our calculations, $\Gamma$ is fixed at $12.8$~meV (this value has already been proven to work well for the analysis of the diagonal conductivity spectrum $\sigma_{xx}(\omega)$ \cite{bitei_sigmaxx}).

The carrier densities $n_c$ of different samples were first estimated from the plasma edge of the free carriers in the \textit{ab}-plane (depicted in the inset of Fig.~1(b)) using $\sqrt{n_ce^2/(m^*\epsilon_0\epsilon_{\infty})}$, where $m^*$ is the effective carrier mass, and $\epsilon_{\infty}$ the high-frequency (background) dielectric constant as determined by the ARPES and previous optical measurements, respectively \cite{bitei_arpes,bitei_sigmaxx}. The effect of carrier doping may be treated within the rigid band approximation and then the carrier concentration dependence of the Fermi energy shows a monotonic behavior as plotted in Fig.~1(d).
To study the effect of carrier doping on MO response, we targeted the samples with representative carrier concentrations, whose $E_F$ are above (sample \#1 and \#2), around (\#3), and below (\#4) the Dirac point.
The high-quality single crystal was grown by the chemical vapor transport method and $n_c$ was controlled by the growth conditions for samples \#1 and \#3. On the other hand, the Bridgman method has been employed as described in \cite{bitei_arpes} for samples \#2 and \#4, and $n_c$ has been tuned by doping Sb and Cu, respectively. As indicated in Fig.~1(b) the corresponding carrier densities estimated from the plasma frequencies are $n_c\simeq$ 7.5, 5.4, 3.4, and 0.9$\times10^{19}/$cm$^3$ for samples \#1--4, respectively.
Another way to determine the carrier concentration would be via Hall measurements. Importantly, however, in the case of BiTeI the relation
$R_H=-1/(n_ce)$ does not hold. Figure~1(c) shows the calculated dc Hall response as a function of $n_c$ with and without taking into account SOI, which clearly demonstrates that the conventional Hall data analysis under(over)estimates the carrier density above (below) the Dirac point.
The origin of the large enhancement of $B/\rho_{xy}$ (or the suppression of $R_H$) around the Dirac point is due to the coexistence of electron and hole pockets, or equivalently to the donuts-like Fermi surfaces in the RSS band.
In the case of samples \#1--3 the results of magnetotransport experiments
in conjunction with the calculations can give the carrier density values $n_c$ in accord with the values estimated from the plasma edge (Fig.~1(b)).
However, this is not the case for sample \#4, where the complex characteristics of the Fermi surface makes the simple estimate based on the plasma frequency less adequate. Consequently, in the following analysis we use $n_c\simeq0.06\times10^{19}/$cm$^3$ as the carrier density of sample \#4 estimated from the realistic band structure model.

\begin{figure}[ht!]
\includegraphics[width=3.4in]{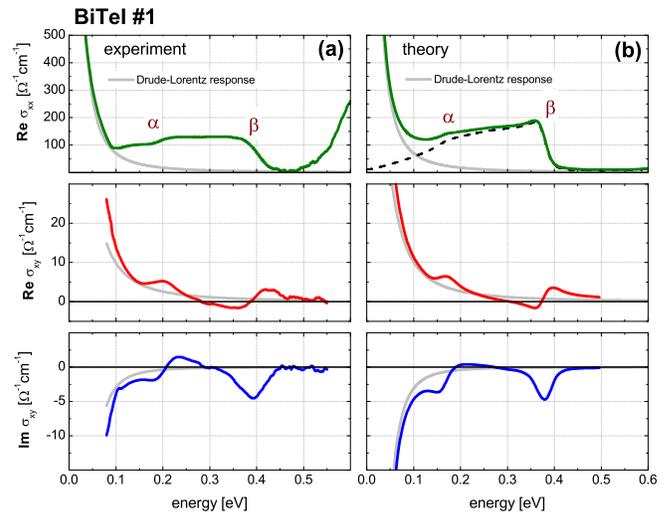}
\caption{(Color online) Comparison between the optical conductivity spectra obtained experimentally (left panels, (a)) and theoretically (right panels, (b)) for sample \#1. Colored lines represent the actual curves, while gray color illustrates the Drude-Lorentz response based on the fit to the low energy part of the experimental $\sigma_{xx}$ (see the text for details). The calculation of $\sigma_{xx}$ (dashed line) does not include the contribution of free carrier excitations, so this part has been added manually by the fit for easy comparison.}
\end{figure}
The experimental spectra for the $ab$-plane of sample \#1 (with $E_F$
well above the Dirac point) are plotted in Fig.~2.
Our theoretical results, also shown in Fig.~2, closely reproduce the experimentally observed spectra both in magnitude and energy position, including $\sigma_{xx}(\omega)$ as well as the real and imaginary parts of $\sigma_{xy}(\omega)$.
Two distinct features in the MO response are discerned in the $\sigma_{xy}(\omega)$ spectra around 0.2~eV and 0.4~eV,
in addition to a Drude-Lorentz response dominating the region below 0.1~eV.
These two resonance structures well correspond to the intraband transitions $\alpha$ and $\beta$ (see Fig.~1(a)) assigned in the $\sigma_{xx}(\omega)$ spectra \cite{bitei_sigmaxx}.
The observed MO response (up to $\pm$5\,$\Omega^{-1}$cm$^{-1}$ at 0.1-0.5~eV) at $3$~T is remarkably large for such a non-magnetic system.
For a comparison, the interband contribution to the transverse conductivity
for typical nonmagnetic semiconductors InSb, InAs, Ge, or GaAs at $3$~T is of the order of
0.01-0.3~$\Omega^{-1}$cm$^{-1}$ as calculated from the Faraday rotation \cite{faraday_theory}.
Even in the case of the ferromagnetic (Gd$_{0.95}$Ca$_{0.05}$)$_2$Mo$_2$O$_7$,
the MO signal in the 0.1-1~eV energy range governed by the Mo $4d$ intraband transitions is an
order of magnitude smaller, and only the spin chirality induced contribution present
in Nd$_2$Mo$_2$O$_7$ is comparable with the present system \cite{nmo}.
The observed MO activity at 3~T can well compare with the typical
magnitude of the MO response ($\sigma_{xy}$) coming from the $p-d$ charge-transfer excitations in oxide ferromagnets \cite{nmo, sro}.
In BiTeI, because of RSS of BCBs and TVBs, the optical transitions between the Rashba-split bands are allowed.
Since these transitions have different dipole matrix elements for left and right
circularly polarized light under applied magnetic field, a magneto-optical response results which scales linearly with the strength of SOI~\cite{oppeneer}. Hence, the large magneto-optical response of BiTeI
is a direct consequence of the gigantic bulk RSS in this material.
This is an astonishing effect of the SOI, since the magnetization
of BiTeI at $B=3$~T is merely of the order of 10$^{-4}$~$\mu_B$/Bi \cite{bitei_chi},
several orders of magnitude smaller than the spontaneous magnetization in the ferromagnets mentioned above.

\begin{figure}[t!]
\includegraphics[width=3.4in]{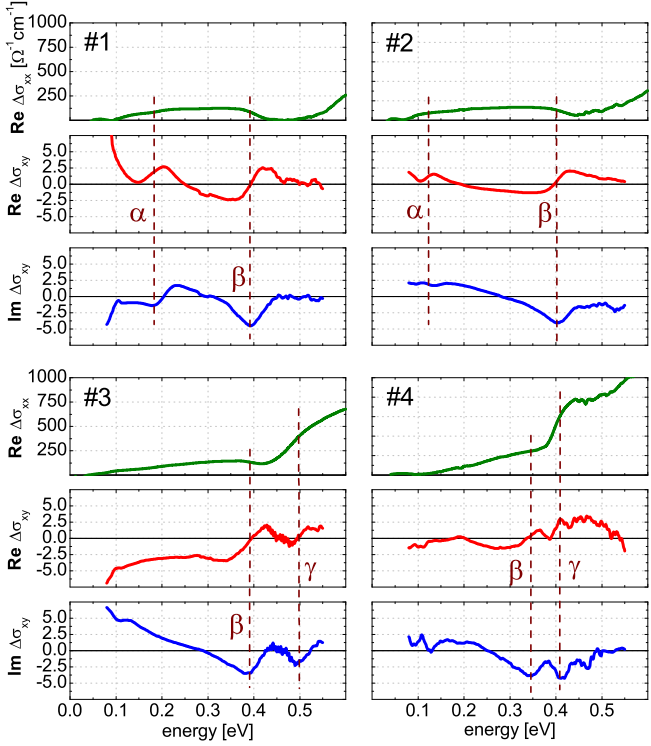}
\caption{(Color online) Systematics of spectral features with changing the Fermi energy $E_F$.
$E_F$ decreases monotonically from sample \#1 towards sample \#4, crossing the Dirac point near sample \#3. Vertical broken lines represent the positions of the structures identified with the intra- ($\alpha$ and $\beta$) and interband ($\gamma$) transitions (see the related discussion).}
\end{figure}
To have a better insight into the consequences of RSS on the MO properties,
the contribution of free carrier excitations has been subtracted by assuming a simple
Drude model expressed
with use of the cyclotron frequency $\omega_c = eB/m^*$, and the relaxation time $\tau = \hbar/\Gamma$.
The dc conductivity $\sigma_{xx}(\omega=0,B=0)$
was determined from the transport experiments, then
$\tau$, the only free parameter, was chosen so as to fit the low energy part of the experimental diagonal conductivity spectra.
The Drude-Lorentz curves are also depicted in Fig.~2.
Figure~3 shows the resulting spectra of $\sigma_{xx}$ and $\sigma_{xy}$,
whose Drude-Lorentz components are subtracted with the similar analysis, for all the
samples investigated. The values of the effective mass ($m^*=0.18m_0$) and the damping constant
obtained from these fits are in good agreement with those found by the ARPES
measurement~\cite{bitei_arpes} as well as those used for the aforementioned theoretical model.

\begin{figure}[t!]
\includegraphics[width=3in]{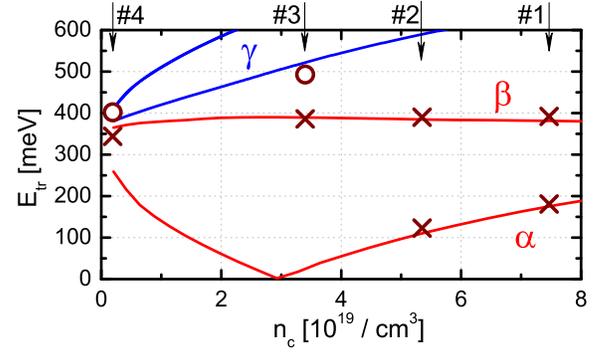}
\caption{(Color online) Observed energies ($E_{tr}$) of intra- and interband transitions in BiTeI with changing carrier density ($n_c$). Solid lines (symbols) represent the calculated (observed) transition energies (energy position of the structures found in the MO spectra). Crosses and open circles correspond to the features identified with the intra- and interband transitions, respectively.}
\end{figure}
Next, we explain the evolution of different spectral features
found in the MO response with varying carrier densities.
Starting with sample \#1, in the $\sigma_{xx}(\omega)$ spectra
one can identify the characteristic features of the intraband transitions as
found previously \cite{bitei_sigmaxx}, namely the vertical transitions to (from) the $E_F$
as depicted by $\alpha$ ($\beta$) in Fig.~1(a).
In external magnetic fields the Zeeman splitting is small compared to RSS, nevertheless the matrix elements of the intraband transitions become different for left and right circularly polarized lights.
The dispersive-like features found in the real part of $\sigma_{xy}(\omega)$ are
the consequence of these processes, giving a clear evidence for the strong SOI in BiTeI.
Lowering the Fermi level ($E_F$) to the level of sample \#2 the $\alpha$ band shifts toward lower energy
with decreasing the magnitude, while the $\beta$ band remains at almost the same position.
Further variation of $E_F$ (as in samples \#3 and \#4) causes slight shifts of $\beta$ as well,
while $\alpha$ is no more discernable in the measured frequency range.
On the other hand, the spectral feature characteristic of the interband transitions $\gamma$ (see Fig.~1(a))
enters into the detectable window from higher energies for samples \#3 and \#4,
observed as a sharp rise in $\sigma_{xx}(\omega)$ at the edge of the interband transition
and as a dispersive spectral shape in the $\sigma_{xy}(\omega)$ spectra.
Figure~4 shows the calculated (solid lines) and observed (symbols) results
for the relevant transition energies as a function of carrier density $n_c$.
The agreement between the experimental results and the theoretical predictions
is excellent in the case of $\alpha$ and $\beta$ transitions (that are assigned to the intraband excitations), and reasonably good in the case of the $\gamma$ (interband) transitions.

To summarize, we have studied the magneto-optical response of BiTeI
with a large Rashba spin splitting
by systematically changing the position of the Fermi level around
the band-crossing (Dirac) point. Given that BiTeI is a non-magnetic system,
the observed MO response arising from the intraband transitions is found to be
huge compared with conventional (spin degenerate) semiconductors.
A theoretical model based on the calculated band structure has been constructed, which
can quantitatively account for the experimental results, and also predict the
significant impact of the Dirac point
on the dc Hall effect.

This research is supported by MEXT Grand-in-Aid No.20740167, 19048008, 19048015,
and 21244053, Strategic International Cooperative Program (Joint Research Type) from
Japan Science and Technology Agency, by the Japan Society for the Promotion of
Science (JSPS) through its ``Funding Program for World-Leading Innovative R\&D
on Science and Technology (FIRST Program)'', the DFG research unit FOR 723,
as well as by Hungarian Research Funds OTKA PD75615, CNK80991, Bolyai 00256/08/11,
and TAMOP-4.2.1/B-09/1/KMR-2010-0002.
G. A. H. S. acknowledges support from MEXT and DAAD.


\begin{thebibliography}{10}
\bibitem{awschalom} D. Awschalom and N. Samarth, Physics {\bf 2}, 50 (2009).
\bibitem{parkin} S. D. Bader and S. S. P. Parkin, Annu. Rev. Cond. Matter Phys. {\bf 1}, 71 (2010).
\bibitem{sinova2} J. Sinova and I. \v{Z}uti\'{c}, Nat. Mater. {\bf 11}, 368 (2012).
\bibitem{winkler} R. Winkler, Springer Tracts in Modern Physics 191 (Springer, Berlin, 2003).

\bibitem{murakami} S. Murakami, N. Nagaosa, and S. C. Zhang, Science {\bf 301}, 1348 (2003).
\bibitem{ganicgev} S. N. Ganichev {\it et al.}, Nature {\bf 417}, 153 (2002).
\bibitem{bennemann} K. H. Bennemann, J. Mag. Mag. Mater. {\bf 200}, 679 (1999).

\bibitem{bitei_arpes} K. Ishizaka et al., Nat. Mater. \textbf{10}, 521 (2011).
\bibitem{bitei_theory} M. S. Bahramy, R. Arita, and N. Nagaosa, Phys. Rev. B \textbf{84}, 041202(R) (2011).
\bibitem{bitei_pressure} M. S. Bahramy, B.-J. Yang, R. Arita, and N. Nagaosa, Nat. Commun. \textbf{3}, 679 (2012).
\bibitem{bitei_sigmaxx} J. S. Lee et al., Phys. Rev. Lett. \textbf{107}, 117401 (2011).

\bibitem{shen} S.-Q. Shen, M. Ma, X. C. Xie, F. C. Zhang, Phys. Rev. Lett. \textbf{92}, 256603 (2004).
\bibitem{burkov} A. A. Burkov and L. Balents, Phys. Rev. B \textbf{69}, 245312 (2004).
\bibitem{xu} W. Xu, C. H. Yang, and J. Zhang, Appl. Phys. Lett. \textbf{91}, 221911 (2007).
\bibitem{kushwaha} M. S. Kushwaha, Phys. Rev. B \textbf{74}, 045304 (2006).

\bibitem{moke} K. Sato, Japan. J. Appl. Phys. \textbf{20}, 2403 (1981).

\bibitem{fukuyama} H. Fukuyama, Prog. Theor. Phys. \textbf{42}, 1284 (1969).
\bibitem{ab-initio}Details of {\it ab-inito} calculations can be found in Ref.~\cite{bitei_theory}.
\bibitem{mahan} G. D. Mahan, \textit{Many-Particle Physics}, 3rd edition (Plenum, New York, 2000).

\bibitem{faraday_theory} L. M. Roth, Phys. Rev. \textbf{133}, A542 (1964).

\bibitem{nmo} I. K\'ezsm\'arki et al., Phys. Rev. B \textbf{72}, 094427 (2005).
\bibitem{sro} Z. Fang et al., Science \textbf{302}, 92 (2003).

\bibitem{oppeneer} P. M. Oppeneer, T. Maurer, J. Sticht, and J. K\"ubler, Phys. Rev. B \textbf{45}, 10924 (1992).
\bibitem{bitei_chi} G. Schober et al., cond-mat/1111.5800, Phys. Rev. Lett. (in press).

\end{thebibliography}
\end{document}